\documentclass[prb,showpacs,nofootinbib,onecolumn,amsmath]{revtex4}
\usepackage{hyperref}
\usepackage{bm,amsfonts,mathtools}
\usepackage{graphicx, wasysym}
\usepackage{textcomp}
\usepackage{amssymb,xcolor,latexsym,epsfig}
\newcommand{\be}{\begin{equation}}
\newcommand{\ee}{\end{equation}}

\usepackage{epstopdf}
\usepackage{hyperref}
\usepackage[utf8]{inputenc}
\usepackage{xcolor}
\def\be{\begin{equation}}
\def\ee{\end{equation}}
\begin{document}
\vspace*{0.5cm}
\title{Very high energy cosmic ray particles from the Kerr black hole at the galaxy center}
\author{Orlando Panella} \email{orlando.panella@infn.pg.it}
\affiliation{Istituto Nazionale di Fisica Nucleare, Sezione di Perugia, Perugia, Italy}
\author{Simone Pacetti} \email{simone.pacetti@unipg.it}
\affiliation{Dipartimento di Fisica e Geologia, Universita' di Perugia, Italy\\
\&\\
INFN Sezione di Perugia, Perugia, Italy}
\author{Giorgio Immirzi} \email{giorgio.immirzi@gmail.com}
\affiliation{Dipartimento di Fisica e Geologia, Universita' di Perugia, Perugia, Italy (Retired)}
\author{Yogendra Srivastava} \email{yogendra.srivastava@gmail.com}
\affiliation{Centro di Ricerche Enrico Fermi, Roma, 00185, Italy\\
\&\\
Emeritus Professor of Physics, Northeastern University, Boston, USA\\
\&\\
INFN, Sezione di Perugia, Perugia, Italy}
\begin{abstract}
\noindent
Conventional general relativity supplies the notion of a vacuum tension and thus a maximum force
$F_{max}=c^4/4G\approx\ 3\times 10^{43}$ Newtons, that is realized for a black hole. In conjunction with the 
Wilson area rule, we are thus led to the surface confinement of the mass of a black hole analogous to the surface 
confinement of quarks. The central result of our paper is that PeV scale protons exist on the surface of a Kerr 
black hole residing at our galactic center that is in concert with the HAWC Collaboration result of a PeVatron at the galactic center.  
\end{abstract}
\keywords{energetic cosmic rays, massive rotating Kerr black holes, perfect mass conductor}
\maketitle

\section{\bf Introduction \label{intro}} 
\noindent
In a seminal paper \cite{Hawking:1975}, Hawking proposed that contrary to the classical notion in which a black hole (BH)
can only absorb particles, a quantum BH can also create and emit particles as if its surface were at a temperature ($T_H$) satisfying 
$k_B T_H= (\hbar \kappa)/(2 \pi)$, where ($c\kappa$) is the {\it surface gravity}
of the BH. For a Schwarzschild BH of mass $M$ and radius $R_s=(2GM/c^2)$, $\kappa=c/(2 R_s)$. Such quantum evaporation
and radiative processes have been studied and confirmed by Bekenstein\cite{Bekenstein:1997} and Unruh\cite{Unruh:1976} among
others. On the other hand, in a series of papers \cite{tHooft1,tHooft2,tHooft3,tHooft4,tHooft5} by 't Hooft and 
co-workers, it has been argued that for a Schwarzschild BH, the radiation temperature should be doubled ($2 T_H$) since the
BH entropy is halved due to a decrease in the number of linearly independent quantum states by precisely the same amount.\\
\\
An approach using two apparently different inputs from the above was employed in \cite{thermo} to obtain the Hawking result 
for the evaporation process of a BH. The first input is that of a {\it maximum gravitational tension} 
\begin{math} F_{max}=\tau= c^4/(4G)\approx 3\times 10^{43} Newtons\end{math} -such a force is only realized at the 
horizon of a BH \cite{Jacobson:1995, Gibbons:2002, Schiller:2004}. The second input employs a gravitational Wilson closed loop action  
to obtain the central result that all matter is confined on the horizon surface by the action-area law via the gravitational tension in the 
closely analogous sense that the Wilson action-area law also describes a surface confinement of quarks in QCD \cite{Wilson:1979}.\\
 Thus, through completely different reasonings, both 't Hooft and reference \cite{thermo}, converged on the basic result that the dynamics of a BH is essentially on its surface -a result substantially different from the {\it standard} 3-dimensional description of it.\\
 \\
However, the detailed formalism \& the tools employed by 't Hooft and that in reference \cite{thermo} being different, they both need to be discussed and understood. For example, in the Abstract of \cite{tHooft5}, it is stated that {\it ...the quantum black hole has no interior, or equivalently, the black hole interior is a quantum clone of the exterior region}. On the other hand, perhaps more prosaically, we consider a BH as a perfect mass conductor with all its mass on its (event horizon) surface. The rather striking results derived in Section(\ref{surface}) from this notion of surface confinement are the central results of the present paper. For example, we are able to show that by virtue of such a surface confinement, the rotating Kerr BH at the galactic center of the Milky Way is driven to become a powerful source of extremely high energy cosmic ray protons. Our calculations in Section(\ref{hess}) quantitatively bear out the recent experimental result from the HAWC Collaboration that calls for the existence of a PeVatron at GC (the galactic center) \cite{Hess:2024}.   

\section{BH maximum tension, Wilson area law \& surface confinement \label{conf}}
\noindent
Conventional general relativity supplies the notion of a vacuum tension $\tau=c^4/(4G)\approx\
 3\times 10^{43}$ Newtons. This vacuum tension determines the maximum force
 (see, ref.\cite{Unruh:1976, Jacobson:1995, Gibbons:2002, Schiller:2004}) $F\leq \tau$ that can be exerted on any material
 body, with equality realized on bodies confined to a horizon surface. This allowed us to prove (rigorously, if any thing is rigorous in relativistic quantum field theory!) the gravitational Wilson action-area result\cite{Wilson:1979} for matter confined on horizons \cite{thermo}. In Euclidean field theory it led us to the well known entropy area theorem on black hole horizons. Recalling our horizon thermodynamics via Eqs.(19-27) from ref.(\cite{thermo}):
 \begin{eqnarray}
\label{t1}
{\rm the\ entropy-area\ relation}:\  (\frac{d\mathcal{S}}{k_B}) = (\frac{\tau}{\hbar c}) (d\mathcal{A});\ {\rm where\ area\ is}: 
\mathcal{A};\nonumber\\
{\rm spherical\ BH\ radius}: R_s=(\frac{2GM}{c^2})= (\frac{\mathcal{E}}{2\tau});\ \mathcal{E}= (Mc^2)\ {\rm is\ the\ BH\ energy};\nonumber\\
{\rm area\ of\ the\ BH\ horizon}\ \mathcal{A}= (4\pi R_s^2)= \pi (\frac{\mathcal{E}}{\tau})^2; {\rm the\ horizon\ entropy}\ \mathcal{S}= \pi k_B (\frac{\mathcal{E}^2}{\hbar c \tau}) ;\nonumber\\
{\rm the\ temperature:}\ (\frac{1}{T})= (\frac{d\mathcal{S}}{d\mathcal{E}});\ \Rightarrow\ k_BT=(\frac{\tau}{2\pi})(\frac{\hbar c}{\mathcal{E}})= (\frac{\hbar c}{4\pi R_s});\nonumber\\
{\rm the\ free\ energy}\ \mathcal{F}= \mathcal{E}- T\mathcal{S}= T \mathcal{S}= \frac{1}{2}\mathcal{E}.
 \end{eqnarray}
 There is surface tension ($\sigma$) of the BH given by
\begin{eqnarray}
\label{st}
\sigma = \big{(}\frac{\mathcal{F}}{\mathcal{A}}\big{)}= \Big{[}\frac{k_B T}{\hbar c}\tau \Big{]}= \Big{[}\frac{\tau}{4\pi R_s} \Big{]}.
\end{eqnarray}
If one draws an equator around the sphere, then the two halves of the sphere attract each other via the vacuum
gravitational tension. Eq.(\ref{st}) is central to our discussion. The surface tension of the horizon is from the
confinement of both energy and entropy on the surface of the black hole. There is no need to discuss what is ``internal'' to the black hole. All of the physical quantities are confined to the horizon surface by the gravitational
tension. A BH with all its mass-energy on the surface is the best analogue to an ideal conductor in electro-dynamics with all its charge on the surface.[Of course, with the most important difference that due to charges of both signs in electro-dynamics, charge neutral objects such as atoms and molecules are not confined to be on the surface of an ideal conductor; clearly, for gravity with positive masses only, all black hole mass is confined to the horizon surface of the BH.]\\  

\section{\bf Practical consequences of all BH mass lying on its surface \label{surface}} 
\noindent
Going beyond ref.(\cite{thermo}), we find that the inter-particle dynamics is dramatically changed as
the BH mass is changed from being distributed  (spherically uniformly) in 3-dimensions 
to being uniformly distributed on the 2-dimensional surface of the BH sphere. \\
\\
Consider the inter-particle distance ($L_2$) in 2 \& ($L_3$) in 3-dimensions for the same radius $R_s$ of the BH
and the same total number ($N_T$) of the particles (here assumed to be nucleons).  
\begin{eqnarray}
\label{p1}
N_T = N_\odot \big{(} \frac{M_{BH}}{M_\odot} \big{)} \approx\ 1.2 \times 10^{57}  \big{(} \frac{M_{BH}}{M_\odot} \big{)};(i)\nonumber\\
L_2 \equiv\ (\frac{Area}{N_T})^{1/2}= (\frac{4\pi R_s^2}{N_T})^{1/2};(ii)\nonumber\\
L_3 \equiv\ (\frac{Volume}{N_T})^{1/3}= (\frac{4\pi R_s^3}{3 N_T})^{1/3};(iii)\nonumber\\
L_3/L_2 = (\frac{N_T}{36 \pi})^{1/6}\approx\ 1.5 \times 10^9  \big{(} \frac{M_{BH}}{M_\odot} \big{)}^{1/6};(iv)\nonumber\\
L_2\approx\ (3\times 10^{-10} Fermi)\times(\frac{M_{BH}}{M_\odot})^{1/2};(v)\nonumber\\
L_3\approx\ (0.45 Fermi)\times(\frac{M_{BH}}{M_\odot})^{2/3};(vi)\nonumber\\
\end{eqnarray}
As expected, the inter particle spacing $L_2$ (when the mass is spread out over the 2-dimensional surface of the BH) compared to $L_3$ (when the BH mass is spread uniformly over a 3-dimensional volume of the BH sphere) is smaller by factors of over a billion; the precise value depending upon how massive the BH is with respect to the solar mass $M_\odot$. We can obtain an estimate of the scale of mean particle energies  ($<E>$)  for a given length scale ($L$) through the uncertainty principle:  
\begin{eqnarray}
\label{p2}
<E>\  \geq [\frac{\hbar c}{L}]
\end{eqnarray}
Some limiting cases as the BH mass varies between say $10 M_\odot$ to a super massive BH of mass $6.5 \times 10^9 M_\odot$ (that has been found by the Event Horizon Telescope to reside at the center of the giant galaxy M87) \cite{EHT:2024}, illustrate the range of variation in the energy scale (and thus the corresponding standard model induced particle dynamics) when all the mass is concentrated at the surface of the BH {\it versus} a 3-dimensional spherically symmetric distribution.\\
There is a factor of 2 difference in the horizon radius between a Schwarzschild and a Kerr BH \cite{LL}. Explicitly, $R_s(Schwarzschild)=(2GM/c^2)= 2 R_s(Kerr)$. As the known massive BH's at the center of various galaxies
are all of the rotating Kerr type, the numerical estimates below are for a Kerr BH: 
\begin{eqnarray}
\label{p3}
<E>_2(M_{BH}=10 M_\odot) \approx\ 0.42 \ EeV ;(i);\nonumber\\
<E>_3(M_{BH}= 10 M_\odot) \approx\ 190\ MeV ;(ii);\nonumber\\
<E>_2(M_{BH}= 4.3 \times 10^6 M_\odot) \approx\ 0.64\ PeV ;(iii);\nonumber\\
<E>_3(M_{BH}= 4.3 \times 10^6 M_\odot) \approx\  33.5\ KeV ;(iv);\nonumber\\
<E>_2(M_{BH}= 6.5 \times 10^9 M_\odot) \approx\ 16.6\ TeV ;(v);\nonumber\\
<E>_3(M_{BH}= 6.5 \times 10^9 M_\odot) \approx\  256.2\ eV ;(vi);\nonumber\\
\end{eqnarray}
These numbers illustrate how different the dynamics near the surface of a BH becomes both because the inter particle dynamics is highly peaked (for our own galactic center Kerr BH in the PeV scale for protons) if there is only a surface mass and also because the heavier the BH the less active it would be energetically. The super massive BH at the center of the M87 galaxy would be almost {\it atomic} were the mass spherically distributed. Clearly, if the surface dynamics effects discussed here are indeed realized, they would provide a clean experimental signal about how the mass is distributed over a BH by virtue of the intensity of particle interactions near the surface of a BH. In particular, we have found -as shown above in Eq.(\ref{p3})- that only if the mass is distributed over the surface of our galactic black hole we shall have  a powerful source of cosmic PeVatron protons.\\
\\
In the following section(\ref{hess}), we shall discuss experimental data from the HAWC Collaboration that distinctly shows that our Kerr BH at the galactic center is a cosmic source of PeV protons thereby confirming our hypothesis that this BH is indeed a perfect mass conductor. Please note that were it much more massive, say the BH had a mass of $6.5 \times 10^9\ M_\odot$ such as the one at the center of galaxy M87, the energy scale for cosmic protons would be a meagre tens of TeV and thus it could not match the experimental UHE gamma ray energy spectrum found between ($6 \div 118$ TeV) \cite{Hess:2024}.\\
An attentive reader may have wondered about the fate of much less massive BH, say with a mass of $10\ M_\odot$, that according to Eq.(\ref{p3}(i)) should be radiating 100 PeV protons. The reader can easily show that such BH's would have a rather short half-life and would not last long. Thus, the mass of BH's that are long lasting but at the same time are also producers of very high energy protons is in a very limited region. We shall return to this interesting subject elsewhere along with a discussion of the critical indices (parameter $\gamma$ in Eq.(\ref{1a})) in the energy spectrum of high energy cosmic ray particles \cite{Widom:2014, Widom:2015a, Widom:2015b, Widom:2015c, Pancheri:2017}.\\          
\section{Experimental verification of our predictions by the HAWC Collaboration \label{hess}}
\noindent
Salient aspects of 7-year data about ultra high energy (UHE) gamma rays  from the galactic center (GC) obtained by the HAWC (High Altitude Water Cherenkov) Observatory \cite{Hess:2024} may be summarized as follows:\\
\begin{itemize}
\item 1. Very high energy  ($6\ TeV \div 118\ TeV$) gamma ray data are best described as originating from a point like source [HAWC J1746-2856] with a power law spectrum for $N$ (i.e., the number current flux density per unit area per unit time) that has been parametrized as
\begin{eqnarray}
\label{1a}
\frac{dN}{dE} = \phi(\frac{E}{26\ TeV})^{\gamma};\nonumber\\
\nonumber\\
\gamma=-2.88\pm 0.15 -0.1;\ \ \phi=1.5\times 10^{-15}\ (TeV cm^2\ sec.)^{-1}\pm 0.3\ (+0.08,-0.13).
\end{eqnarray}
\item 2. There is no evidence of a spectral cut-off \cite{Greisen:1966,ZK:1966} up to 100 TeV in the HAWC data.
\item 3. The HAWC Collaboration concludes that the UHE gamma rays detected by them originate via hadronic interaction of PeV cosmic ray protons with the dense ambient gas and confirms the presence of a proton PeVatron at the GC but they do not provide a mechanism for it.
\end{itemize} 
The experimental results (1-3) require a dynamical mechanism for the existence of a proton Pevatron and it is pleasing to note that such a Pevatron emerges naturally under our central notion in the present paper that the Kerr BH at the center of our galaxy is indeed a perfect mass conductor in that all its mass is concentrated at its event horizon 
[{\it vedi} Eq.(\ref{p3}(iii)].

\section{Conclusions \label{con}}
\noindent
For (super)massive BH that are supposed to be at the center of most galaxies, we have shown that the inter-particle distance can indeed become very small when all mass is distributed over the horizon surface. This implies that the mean energy of the particles can be driven to extremely large values. And we have explored here the notion that cosmic rays are emitted from the surfaces of BH just as they are from neutron stars. The cosmic rays themselves are in 
the stellar atmospheric winds blowing away from the  source. The cosmic rays are equivalently nuclei which are
 evaporating from the surface. 
The method of computing the energy distribution 
of the evaporated cosmic rays is closely analogous to those employed 
by Landau and Fermi for evaporation of nucleons in the Bohr-Mottelson 
liquid drop model \cite{Widom:2014, Widom:2015a, Widom:2015b, Widom:2015c, Pancheri:2017}\\
\\
Our prediction about PeV scale protons at the surface of the Kerr black hole situated at our galactic center agrees with that of the HAWC Collaboration claiming the existence of a PeVatron proton source at the galactic core [through their observation of ultra high energy photons with energies between ($6\div118$) TeV]. In view of the discovery of a PeVatron at the galactic core through the observation of UHE photons tempts us to call super massive BH, Bright Holes rather than Black Holes.\\
\\
It seems inescapable to generalize Wheeler's {\it no hair theorem for a BH} to read: {\it A BH is characterized by its mass; its angular momentum \& its total charge: all lying on its (event horizon) surface}.\\
 \\
 \section{Acknowledgments}
YS gratefully acknowledges some very useful correspondence with Professor G. 't Hooft and he also thanks Professor U. Heintz for a careful reading of the manuscript.\\

\end{document}